\documentclass[usenatbib,useAMS]{mnras}

\usepackage{newtxtext,newtxmath}
\usepackage{amsmath}
\usepackage{bm}

\usepackage[T1]{fontenc}
\usepackage{ae,aecompl}

\usepackage{multicol}
\usepackage{graphicx}
\usepackage{pdflscape}
\usepackage{color}
\usepackage{xcolor}
\usepackage{csvsimple}

\usepackage[normalem]{ulem}

\newcommand{\removed}[1]{}

\newcommand{\codefont}[1]{{\texttt{#1}}}
\newcommand{\Mpch}{\,h^{-1}\mathrm{\,Mpc}}

\title[Anti-halo Void Catalogue]{An Anti-halo Void Catalogue of the Local Super-Volume}

\author[S. Stopyra et al.]{Stephen Stopyra$^{1}$\thanks{Contact e-mail: \href{mailto:stephen.stopyra@fysik.su.se}{stephen.stopyra@fysik.su.se}},
	Hiranya V. Peiris$^{3,1}$, Andrew Pontzen$^{2}$, Jens Jasche$^{1}$, Guilhem Lavaux$^{4}$
	\\
	$^{1}$The Oskar Klein Centre for Cosmoparticle Physics, Department of Physics, Stockholm University, AlbaNova, Stockholm SE-106 91, Sweden\\
	$^{2}$Department of Physics and Astronomy, University College London, Gower Street, London WC1E 6BT, UK\\
	$^{3}$ Institute of Astronomy and Kavli Institute for Cosmology, University of Cambridge, Madingley Road Cambridge, CB3 0HA, United Kingdom\\
	$^{4}$ Sorbonne Université, CNRS, UMR 7095, Institut d’Astrophysique de Paris, 75014 Paris, France
}

\date{Accepted XXX. Received YYY; in original form ZZZ}
\pubyear{2022}

\begin{document}
\label{firstpage}
\pagerange{\pageref{firstpage}--\pageref{lastpage}}
 \maketitle

	\begin{abstract}
		We construct an anti-halo void catalogue of $150$ voids with radii $R > 10\Mpch$ in the Local Super-Volume ($<135\Mpch$ from the Milky Way), using posterior resimulation of initial conditions inferred by field-level inference with Bayesian Origin Reconstruction from Galaxies (\codefont{BORG}). We describe and make use of a new algorithm for creating a single, unified void catalogue by combining different samples from the posterior. The catalogue is complete out to $135\Mpch$, with void abundances matching theoretical predictions. Finally, we compute stacked density profiles of those voids which are reliably identified across posterior samples, and show that these are compatible with $\Lambda$CDM expectations once environmental selection (e.g., the estimated $\sim 4\%$ under-density of the Local Super-Volume) is accounted for. 
	\end{abstract}


	\section{Introduction}
	\label{sec:introduction}
	
	\begin{figure*}
		\centering
		\includegraphics[width=\textwidth]{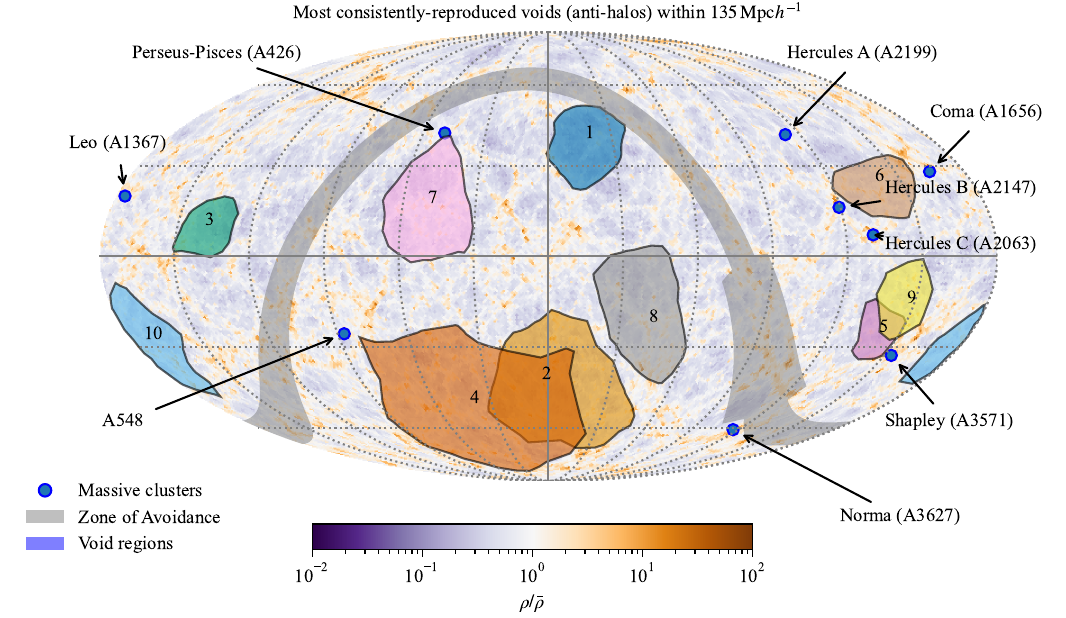}
		\caption{\label{fig:skyplot} Projected locations of the 10 anti-halo voids which appear most consistently across 20 MCMC samples (all have reproducibility scores of $0.95$ or above). Numbers follow the same ordering as in Table~\ref{tab:anti-halos}, sorted in descending order of reproducibility score. Voids 1-6 appear in all 20 posterior resimulations, while voids 7-10 are each missing in only 1 of 20 samples. Note that apparent void sizes are driven by distance-dependent projection effects.}
	\end{figure*}
	
	Cosmic voids -- regions of the Universe with significantly lower density than the clusters and filaments where most galaxies are found -- make up the majority of the Universe's volume, and provide a pristine environment for probing cosmology and fundamental physics~\citep{1981ApJ...248L..57K,zeldovich1982giant}. In particular, the shapes of voids, via their ellipticity distribution~\citep{PhysRevLett.98.081301} or the Alcock-Paczynski test~\citep{1979Natur.281..358A}, and their abundance via the void size function, can be used to constrain cosmological parameters~\citep{Sutter_2012,Sutter:2014oca, contarini2023cosmological}. Furthermore, voids can probe modifications to General Relativity on large scales which may be hidden by screening in denser areas of the Universe~\citep{spolyar2013topology,joyce2015beyond}, and probe the dark energy equation of state~\citep{lee2009constraining}.
	
	However, there are a large number of different void definitions in use. A commonly-used empirical approach is to identify voids from the morphology of the density or galaxy field via the watershed algorithm, using codes such as \codefont{ZOBOV}~\citep{Neyrinck:2007gy} or \codefont{VIDE}~\citep{Sutter:2014haa}. Other approaches include spherical voids~\citep{10.1111/j.1365-2966.2005.09500.x,2006MNRAS.373.1440C,10.1093/mnras/stv019}, and the related popcorn voids~\citep{paz2023guess}. Most of these definitions use the present day galaxy distribution, which  traces the underlying dark matter, and the morphology of the cosmic web. This makes the relationship between cosmological initial conditions and the shapes, sizes, and abundance of voids challenging to model, in contrast with halos where the relationship can be understood via excursion set models~\citep{1991ApJ...379..440B} and modifications thereof. Several authors have studied the abundance of voids with similar excursion-set approaches~\citep{Sheth:2003py,Jennings:2013nsa}, but it has proven challenging to relate these calculations to observational data ~\citep{Nadathur:2015qua}. 
 
 One approach in which the abundance can be directly predicted, however, is the \emph{anti-halo} model of voids, proposed by~\citet{Pontzen:2015eoh}, and further studied by~\citet{stopyra2021build,shim2021identification,shim2023cluster,desmond2022catalogues}. In the anti-halo approach, voids in an $N$-body simulation are defined in analogy to clusters by performing an `anti-universe' simulation with density-inverted initial conditions: by exchanging under-dense and over-dense regions in the initial conditions, particles identified as belonging to halos in the anti-universe simulation correspond to voids in the original simulation. The abundance of such voids is then directly related to the initial conditions via the (anti-)halo mass function, placing halos and anti-halo voids on equal footing.
	
	Because anti-halos are defined with reference to structure formation theory, identifying them in observational data requires additional steps. ~\citet{shim2023cluster} used an approach with Gaussian smoothing to infer the locations of anti-halos from galaxy data, calibrated using simulations and mock galaxy catalogues. To construct a anti-halo catalogue directly from the dark matter density field, one would require access to the initial conditions of an $N$-body simulation which, when evolved forward, corresponds to the actual density field of the present-day Universe. Field-level inference is an ideal approach for achieving this, since it can be used to sample the posterior distribution of possible initial conditions conditioned on the observed galaxy distribution. The Bayesian Origin Reconstruction from Galaxies (\codefont{BORG}) algorithm~\citep{jasche2019physical} is one such example of field-level inference. By utilising the \emph{posterior resimulation} technique~\citep[hereafter S23]{stopyra2023towards}, i.e., resimulating samples and density-reversed samples from the posterior distribution, and identifying anti-halo voids in each sample, it is possible to build an ensemble of samples from the posterior distribution of anti-halo catalogues. 
 
 Recently,~\citet{desmond2022catalogues} built an anti-halo void catalogue based on initial conditions derived from the Markov chain produced by~\citet{jasche2019physical}. However, it was subsequently demonstrated by S23 that the forward model used for field-level inference by~\citet{jasche2019physical} was insufficiently accurate to describe massive halos (and anti-halos). This led to an overestimate of anti-halo masses and therefore, an overestimate of the anti-halo mass function (see S23).
	
	In this work, we present an anti-halo catalogue for voids in the Local Super-Volume using posterior resimulations of initial conditions from a new Markov chain computed by S23, based on Bayesian Origin Reconstruction from Galaxies  \citep[\codefont{BORG},][]{2013MNRAS.432..894J} applied to the 2M++ galaxy catalogue~\citep{2mppPaper}. This chain uses a 20-step \codefont{COLA}~\citep{tassev2013solving} forward model to increase the accuracy of initial condition inference relative to earlier work by \citet{jasche2019physical}, eliminating the aforementioned spurious excess of anti-halos seen by~\citet{desmond2022catalogues}. We also introduce an algorithm for combining lists of anti-halos from different posterior resimulations into a combined catalogue of well-constrained voids, robustly identified across the different posterior samples.

	The structure of this paper is as follows. In Sec.~\ref{sec:methods} we explain how field-level inference with \codefont{BORG} can be used to produce an ensemble of anti-halo lists via resimulation of posterior samples, and then outline our algorithm for combining these lists into a single, unified void catalogue. In Sec.~\ref{sec:results}, we compare the abundance of voids in the combined catalogue to $\Lambda$CDM predictions using the anti-halo mass function and void size function, and also compute the stacked void density profile. We discuss the implications of these results for the compatibility of the Local Super-Volume with $\Lambda$CDM in Sec.~\ref{sec:discussion}.

	\section{Methods}
	\label{sec:methods}

	In this work we make use of the technique of posterior resimulation (see S23) in order to link the anti-halo void definition with data, and thereby construct a void catalogue from the 2M++ galaxy survey~\citep{2mppPaper}. We begin by briefly outlining the posterior resimulation technique and describing how we selected $N=20$ posterior samples for resimulation in Sec.~\ref{subsec:posterior_resim}. Using methods outlined in Sec.~\ref{subsec:antihalos}, we use these posterior resimulations to construct $20$ lists of anti-halos, one for each posterior sample. In Sec.~\ref{sec:construction} we show how to combine these samples into a single catalogue of  robustly-identified voids. We then describe how to construct the stacked void profile of this combined catalogue in Sec.~\ref{sec:void_profiles}.
	
	\begin{figure*}
		\centering
		\includegraphics[width=\textwidth]{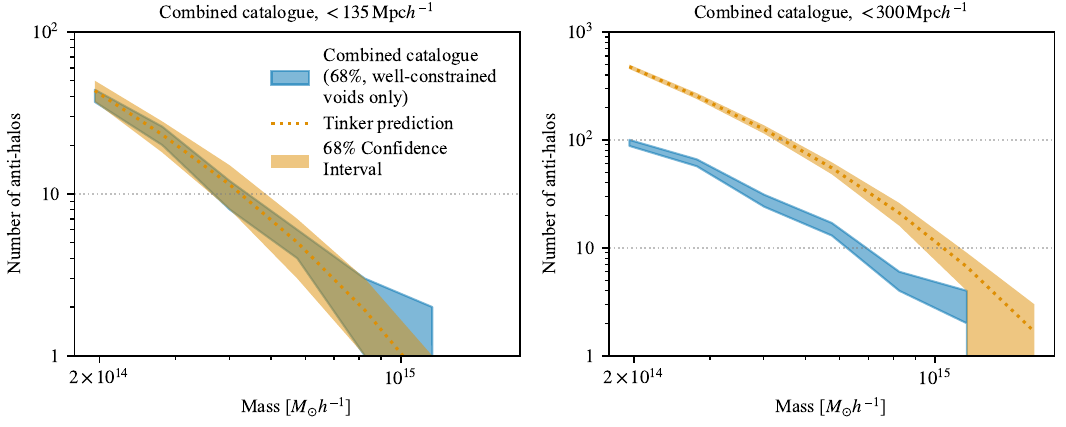}
		\caption{\label{fig:mass_functions} Mass functions for the combined catalogue of high-confidence voids. Left: voids within $135\Mpch$. The signal-to-noise for these is high, and the catalogue is complete for the most massive anti-halos (above $4\times 10^{15}M_{\odot}h^{-1}$), matching the expected number from $\Lambda$CDM predictions. Low mass, low-radius voids tend to be more prior-driven than higher mass voids, and the algorithm therefore excludes them at a higher rate. Right: all voids within $300\Mpch$. This catalogue is largely incomplete, since there are not many more voids at this distance which can be identified with sufficient signal-to-noise to rule out spurious alignment between MCMC realisations. We estimate the bin count and its error using a bootstrap approach (see Sec.~\ref{sec:amf}).}
	\end{figure*}

 	\subsection{Posterior resimulation with BORG}
 	\label{subsec:posterior_resim}	

    The central technique in the present work will be {\it posterior resimulation}, as presented in our recent work, S23. The approach uses field-level inference to draw samples from the posterior distribution of possible initial conditions compatible with data (in this case, the 2M++ galaxy catalogue of the Local Super-Volume). These initial conditions can then be evolved to redshift $z=0$ using an $N$-body solver. We can also evolve a `reversed' set of initial conditions, in which the density contrast is inverted (swapping under-densities and over-densities); this allows anti-halo catalogues to be generated for each sample. 
    
    The first crucial ingredient in reliable posterior resimulation is that the accuracy of the field-level inference is sufficient to accomplish the science goals. For our present purposes, it is vital to reproduce the halo and anti-halo mass function accurately within resimulations. S23 studied the required accuracy in the gravity solver which, during the inference process, is by necessity approximate; we found that it is possible to balance computational speed with physical accuracy, even to the point of obtaining reliable mass estimates of individual clusters. The same level of accuracy was shown to be sufficient for voids, we therefore use results from the same Markov chain, obtained with a 20-step \codefont{COLA} gravity solver, in the present work. The inference makes use of the~\citet{neyrinck2014halo} bias model, which describes galaxy bias as a power-law with an exponential cutoff in order to allow for different behaviour in underdense regions. See S23 for further details on why an approximate gravity solver can lead to convergence on cluster and void masses via posterior resimulation.  
    
    A second crucial ingredient is that the structures being resimulated (clusters, or voids) are reliably represented across posterior samples, more frequently than is found by cross-matching objects in independent, random simulations. This is because the posterior contains a mix of prior-driven and data-constrained structures. To reliably identify the latter, it is necessary to establish that their frequency of occurrence in posterior samples is significantly more likely than by chance alone. We discuss this requirement and its implementation in practice in Sec.~\ref{sec:construction}.
        
    The $20$ samples used in this work were drawn from the MCMC chain computed by S23. Balancing available computational resources and the fact that the MCMC samples are correlated, we picked a subsample for resimulation which saturates the available statistical power in constraining structures in the relevant mass range. The samples were chosen from the portion of the chain after burn in, with a spacing of $300$ samples between every sample which was resimulated. This spacing was chosen since it was longer than the correlation lengths of all the relevant model parameters of the field-level inference, as well as the correlation lengths of the density field around the largest clusters (see S23 for further details). We further investigated the effect of doubling the number of samples from 10 to 20 on the catalogue, finding that this did not lead to significant changes in the catalogue. We now describe the use of these 20 samples for posterior resimulation.
    	
	The posterior resimulations contain $512^3$ particles over a $677.7\Mpch$ box, evolved to $z=0$ from the inferred initial conditions with \codefont{GADGET2}~\citep{Springel:2005mi}. While \codefont{BORG} produces a $256^3$ output, we interpolate this to $512^3$ for resimulation so that shot noise can be suppressed for the largest halos. Density-reversed initial conditions are created using \codefont{genetIC} \citep{stopyra2021genetic} and also evolved to $z = 0$ to provide reverse (anti-Universe) simulations for generating anti-halos. We also run 20 independent random simulations (with both forward and reverse initial conditions for each) with the same settings, but random initial conditions, to use as a $\Lambda$CDM reference. Random initial conditions are generated on a $256^3$ grid and interpolated up to a $512^3$ grid with \codefont{genetIC} in order to replicate the procedure used for posterior resimulation and account for any differences arising from the interpolation step. Throughout this work, we use the Planck 2018 cosmological parameters~\citep{aghanim2020planck}, with lensing and baryon acoustic oscillations: $\Omega_m = 0.3111,\sigma_8 = 0.8102, H_0 = 67.66\mathrm{km}\,\mathrm{s}^{-1}\,\mathrm{Mpc}^{-1}, n_s = 0.9665, \Omega_b=0.049$.
	
	\begin{figure}
		\centering
		\includegraphics[width=0.45\textwidth]{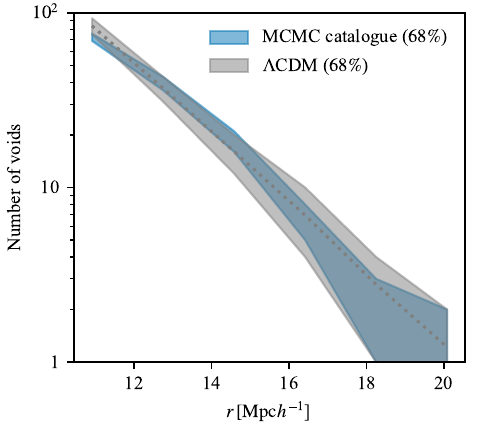}
		\caption{\label{fig:vsf} Void size function for the catalogue compared with the $\Lambda$CDM expectation, showing consistency across all mass bins and demonstrating the completeness of the catalogue for $>10\Mpch$ voids within $135\Mpch$ of the Milky Way. As in Fig.~\ref{fig:mass_functions}, we compute a weighted histogram using the scheme outlined in Sec.~\ref{sec:amf}.}
	\end{figure}
	
	\subsection{Identifying anti-halos in posterior resimulations}
	\label{subsec:antihalos}
	
	To identify anti-halos from posterior resimulations, we apply the halo finder \codefont{AHF}~\citep{AHF} to the reverse simulations. The particles corresponding to each halo in the reverse simulation are then mapped to the corresponding particles in the forward simulation, where they define the anti-halos. We compute a Voronoi tessellation using the \codefont{VOBOZ} code~\citep{neyrinck2005voboz} in order to assign every particle $i$ in the anti-halo a volume, $V_i$. We then compute for each anti-halo the volume-weighted centre, $\mathbf{x}_{\mathrm{VW}}$, and effective radius, $r_{\mathrm{eff}}$, given by
	\begin{align}
		\mathbf{x}_{\mathrm{VW}} =& \frac{1}{V}\sum_{i}V_i\mathrm{x_i},\label{eq:VWcentre}\\
		r_{\mathrm{eff}} =& \left(\frac{3V}{4\pi}\right)^{1/3},\label{eq:Reff}
	\end{align}
	where $i$ runs over all particles in the anti-halo and $V=\sum_iV_i$ is the total volume of all Voronoi cells. The impact of redshift space distortions is already taken into account during the inference step in Sec.~\ref{subsec:posterior_resim}. We refer the reader to S23 for more details, but in brief the gravity solver is used to model velocities at each point, which allows the density field to be computed in redshift space for comparison with galaxy catalogue data. Applied to each resimulation, this gives us $N=20$ separate lists of anti-halos (independent samples from the posterior distribution), which we now combine to create a single robust anti-halo catalogue.
	
	\begin{figure*}
		\centering
		\includegraphics[width=\textwidth]{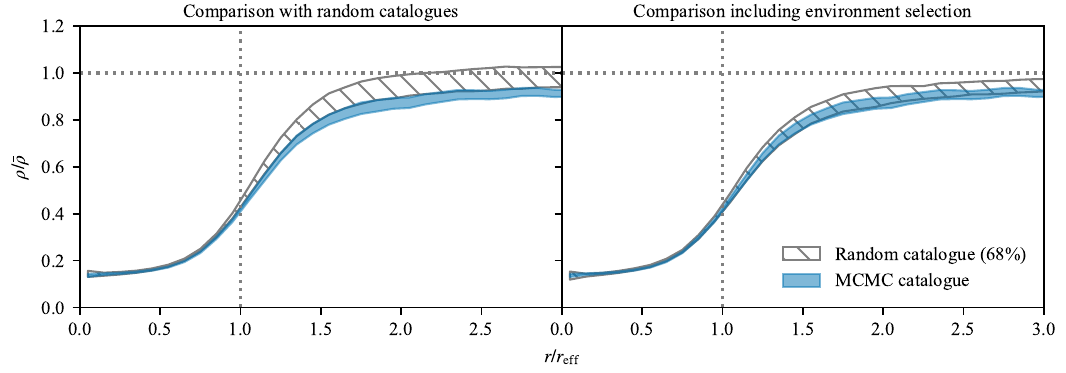}
		\caption{\label{fig:profiles} Density profile for the catalogue within $135\Mpch$ (blue) compared with the $68\%$ interval for the distribution of profiles in randomly-selected $135\Mpch$ regions assuming $\Lambda$CDM (grey solid lines). Without accounting for environmental effects (left) the density between 1 and 3 effective radii in the posterior catalogue is low at the $1\sigma$ level. However, if we select regions from $\Lambda$CDM simulations which agree with the $\sim 4\%$ under-density of the Local Super-Volume (right), and additionally account for the distribution of central and average densities (see Sec.~\ref{sec:void_profiles}) the profile is consistent with $\Lambda$CDM expectations.}
	\end{figure*}
	
	\subsection{Construction of the anti-halo catalogue}
	\label{sec:construction}

    The constraining power of the galaxy catalogue is variable within the volume that it samples, due to signal-to-noise fluctuations in the data. Therefore some anti-halos identified in the previous step are expected to be well-constrained by data and stable across the $N=20$ independent posterior resimulations, while others will not be reproduced across the  samples. In addition, even those which are present in each sample will differ in varying degrees as to their exact locations and radii. Therefore, creating a combined catalogue requires us to associate these slightly differing anti-halos found in different resimulations with a single, reliably reproducible void. A method for identifying candidates for the ``same'' halo between resimulated posterior samples has been proposed by~\citet{hutt2022effect} and recently~\citet{2023arXiv231020672S} have proposed an approach based on the overlap of Lagrangian regions. However, for reliable posterior resimulation it is necessary to assess how well-constrained an individual cluster or void is by the data, as opposed to being largely prior driven. This can be quantified by defining a \emph{reproducibility score}, defined as the fraction of posterior samples in which a void appears, and comparing with independent random catalogues to assess whether a match is significantly more likely than chance.
    
    Our primary desiderata for creating a combined anti-halo catalogue are:
    \begin{enumerate}
    \item The catalogue should prioritise reliability and reproducibility over completeness;
    \item Each posterior sample should be on an equal footing, i.e., anti-halos identified in any one resimulation should not be privileged over those from another resimulation;
    \item Those anti-halos appearing in the combined catalogue should closely agree in their centres and radii across a high fraction of the resimulations.
    \end{enumerate}
	
In order to address desideratum (i), we filter the anti-halos identified in each resimulation based on thresholds for signal-to-noise and minimum radius corresponding to the resolution limit of the simulations.
We only seek to include voids in the final catalogue with radii $r_{\mathrm{eff}}$ above $10\,\Mpch$, which corresponds approximately to the minimum mass at which halos can be resolved with least 100 particles in posterior resimulations. However, to avoid edge effects, we perform the matching with smaller voids down to $5\Mpch$, and retain only voids with mean radii above $R > 10\Mpch$ after matching. In order to apply the signal-to-noise cut, we remove anti-halos for which the mean value of $(\delta/\sigma_{\delta})^2$ over the voxels within a sphere of radius $r_{\mathrm{eff}}$ about an anti-halo's centre is less than $10$, where $\delta$ is the mean density contrast over all MCMC samples in a voxel, and $\sigma_{\delta}$ the standard deviation of the density contrast over a voxel.

The above procedure yields $20$ filtered lists of anti-halos which we pass on to an iterative matching procedure as follows. We start by initializing a candidate list of voids using the centres and effective radii from one of the resimulations, which we denote as resimulation $1$. We then match these candidates with the anti-halos in resimulation $2$ by demanding that the centre and radii are equal within a specified threshold. This introduces two parameters which reflect the tolerance on the location and radius matches respectively. We call these tolerance parameters $\mu_S$ and $\mu_R$ for the centre and radius respectively. 

In practice, these must scale with the size of the voids being matched, and we therefore express them as dimensionless fractions of the radius. Since the anti-halo radius may differ from the radius of the candidate void, the search radius tolerance, $\mu_S$, is defined as the ratio of the maximum permitted distance between the centres of the voids to the geometric mean between their radii. The radius ratio is defined as the \emph{minimum} allowed ratio between the lowest radius and highest radius void in a pair. Explicitly, consider a pair of voids with centres $\mathbf{x_1},\mathbf{x_2}$, and radii $R_1,R_2$ respectively. Assuming that $R_1 < R_2$ without loss of generality, the radius ratio and search distance ratio must satisfy
\begin{align}
	\mu_R \leq & \frac{R_1}{R_2} \leq 1, \\
	\mu_S \geq & \frac{|\mathbf{x}_1 - \mathbf{x}_2|}{\sqrt{R_1R_2}} \geq 0.
\end{align}

If more than one anti-halo in resimulation $2$ satisfies our criteria for a given candidate void, we flag the match as ambiguous and treat it as though no match were found. Otherwise, the matching procedure is repeated starting from the anti-halo list from resimulation $2$ and matching it back to the candidate void list. Any anti-halos where this backwards map results in a different identification are again flagged as ambiguous, and treated as though there was no match at all. In summary, a candidate void is only treated as identified in resimulation $2$ if there is a unique two-way match onto an anti-halo, addressing desideratum (iii).

Once the matching of the candidate void list onto resimulation $2$ is complete, the procedure begins again with resimulation $3$, and so on. At this point, one may calculate the reproducibility score for each candidate void, reflecting how many of the $N$ independent resimulations it appears within. Candidate voids with a high reproducibility score are more robust than those with a low reproducibility score, and we will shortly return to how we threshold the candidate voids on this robustness measure.

However, at this point in the algorithm, the candidate void list has unfairly privileged resimulation $1$ over the other $N-1=19$ resimulations, in contradiction with our desideratum (ii). There is therefore the need to iterate the candidate void list, replacing its initial centres and radii with the mean over all anti-halos counted as a match. We find that typical candidate voids converge to a single shared centre and radius within a few iterations, but any which do not converge within 100 iterations are discarded as unreliable\footnote{In practice, this did not affect any of the voids above $10\Mpch$ in our experiments, but we retain this stop condition so that the algorithm will not become stuck if handling less well-constrained and/or smaller voids.}.

	In order to choose the values of $\mu_R$ and $\mu_S$ in practice, we made use of a visually-confirmed `curated set' of voids which are found with high confidence in a high fraction of the void lists obtained from the $20$ posterior resimulations. The curated set of voids consisted of 30 voids selected by hand which were  verified by visual inspection to have corresponding voids in more than half of samples. We inspected the spread of centres and radii of these curated voids and chose $\mu_R = 0.75, \mu_S = 0.5$ as accurately identifying unique matches for this curated void set while yielding as tight a tolerance as possible on anti-halo centres and radii across the resimulations. This value of $\mu_R$ corresponds to allowing radii to differ by no more than $25\%$ with respect to the larger radius in a pair, while the chosen value of $\mu_S$ means that void centres can differ from each other by no more than half the geometric mean of their radii. We explored the parameter space of possible values around this canonical set of parameters to confirm that this choice recovered the curated voids while also producing a final catalogue with high completeness. 

	After this stage, it is necessary to impose a reproducibility score cut in order to remove poorly-constrained voids from the catalogue. This cut must be carefully chosen in a radius-dependent manner in order to obtain a pure catalogue. We carried out the following procedure to minimize the possibility of spurious voids making it into the combined catalogue via chance alignments. First, we applied the algorithm specified above to a set of anti-halo lists created from $20$ independent, random simulations. For both this random catalogue and the combined catalogue obtained from the posterior resimulations, we then computed the reproducibility score of each void. We observed that the threshold for a significant reproducibility score was radius dependent: smaller voids would frequently be matched by chance in random catalogues, while this was rarer for larger voids. To allow for this, we binned both catalogues in seven bins by the void mean radius, equally spaced between $10\mathrm{-}20\Mpch$, and removed all voids in the combined catalogue whose reproducibility score was lower than the 99th percentile for the same bin in the random catalogue. These bins were chosen since they subdivide the radius range as much as possible, while also yielding at least $\sim O(1)$ voids in even the highest radius bins. We checked that changing the binning scheme does not significantly affect the catalogue as long as more than three radius bins are used. The reproducibility score for each void is given in Table~\ref{tab:anti-halos}. The threshold reproducibility scores as a function of radial bin are given in Table~\ref{tab:thresholds}. Overall, the reproducibility score cut removes $44\%$ of the voids above $10\Mpch$ in the combined catalogue obtained in the previous step.
 
	We use this filtered set as our final catalogue. We estimate the properties of each void in the final catalogue -- such as the mean centre, effective radius, and central density -- as the mean of that property over the corresponding anti-halos in each posterior resimulation in which it is found. We provide the reproducibility score and signal-to-noise of each void as a measure of the void's reliability. 

    \newpage
	
	\subsection{Computation of void density profiles}
	\label{sec:void_profiles}
	
	The void dark matter density profile as a function for radius is sensitive to changes in the cosmology~\citep{dai2015void,chantavat2016cosmological}, deviations from the standard laws of gravity in low density regions~\citep{falck2018using}, and the effects of neutrinos~\citep{contarini2021cosmic,Schuster_2019}. It is, however, not directly observable, and typically the galaxy number density profile is used instead~\citep{hamaus2014universal,Nadathur:2014qja,2023JCAP...05..031S}. An innovation in our work is that the  dark matter distribution has been inferred and can be used to construct a stacked (i.e. averaged) dark matter density profile from our combined final catalogue, as described below. 
 
	We first estimate the density profile for each void by averaging across the anti-halos matched to it within the posterior resimulations. Specifically, we construct spherical shells around the volume-weighted centre in Eq.~(\ref{eq:VWcentre}) for each MCMC sample and use the total mass of dark matter particles in each to obtain a density estimate using a volume-weighted mean, as outlined by ~\citet{Nadathur:2014qja}, but taken over the $N$ realisations of the \emph{same} void in all samples. The uncertainty on each individual void profile (with index $i$) is calculated by computing the variance, $\sigma_{i}^2(r)$, of this profile over all samples. These density estimates are then averaged between the individual anti-halos to obtain a final estimate for the void density profile as a function of the scaled radius, $r/r_{\mathrm{eff}}$.
	
	We then stack these posterior profiles across the entire catalogue, again following the volume-weighted stacking procedure of~\citet{Nadathur:2014qja}. Since we use dark matter particles rather than galaxy tracers, Poisson errors are subdominant to profile variability and we therefore use a different procedure to that used by~\citet{Nadathur:2014qja} to compute the uncertainty in the stacked profile. This procedure is outlined in~\citet{stopyra2021build} and gives the  variance, $\sigma_{\mathrm{mean}}^2(r)$, of the volume-weighted mean profile in a stack, $\rho_{\mathrm{mean}}(r)$, as
	\begin{equation}
		\sigma_{\mathrm{mean}}^2(r) = \left(\sum_{i}w_i^2\right)\sigma^2(r),
	\end{equation}
	where we sum over all voids, $i$, in the stack with volume-weights $w_i=V_i\sum_{i}V_i$ for void volume $V_i$, and $\sigma^2(r)=\mathrm{var}(\{\rho_1(r),\ldots ,\rho_{N}(r)\})$ is the variance of all profiles in the stack and thus common to all voids. This generalises the usual variance of the mean for weighted means, and reduces to $\sigma^2(r)/N$ for the case of $N$ equal weights.
	
	However, because each void in the stack now has a significant uncertainty associated with it, quantified by the variability of the profile of this specific void across the 20 MCMC samples, the procedure requires a slight modification. We add the variance of the profile mean with the posterior variances of each void in the stack, $\sigma_{i}^2(r)$, to obtain
	\begin{equation}
		\sigma_{\mathrm{profile}}^2(r) = \sum_{i}w_i^2\left(\sigma^2(r) + \sigma_{i}^2(r)\right).\label{eq:sigmaProfile}
	\end{equation}
	Eq.~(\ref{eq:sigmaProfile}) then gives the final error bar on the stacked profile, in each radial bin.		  	
	
	To obtain a $\Lambda$CDM profile to compare with, we apply the same stacking procedure to anti-halos within the random simulations previously described (though in this case there is no posterior variance on each individual void profile). In order to construct a fair comparison, we need to account for several environmental selection effects related to the properties of the Local Super-Volume. The Local Super-Volume voids exist within an underdense region of the Universe (the mean density contrast over all MCMC samples is $\delta = -0.043\pm 0.001$), and their central and average void density distributions (defined as the density within 1/4 of the void radius, and the total mass divided by the total volume of the void, respectively) form subsets of void properties that occur within $\Lambda$CDM simulations sampling all environments. To account for these selection effects, our comparison profiles are constructed from anti-halos found within $135\,\Mpch$ spheres with a density contrast matched to that of the Local Super-Volume (in this case, choosing regions whose density contrast lies within a $68\%$ confidence interval of the maximum a posteriori (MAP) estimate, $\delta_{\mathrm{MAP}} = -0.041$\footnote{The difference between the MAP estimate and the mean density contrast is due to the fact that the posterior on the Local Super-Volume density is non-Gaussian, and skewed.}). We randomly select voids from these matched regions in a way that (up to the limits of sample size within each region) reproduces as closely as possible the central/average density distribution of the combined catalogue. To avoid duplication, we retain only spherical regions that do not overlap, giving a total of $144$ density-matched regions over 20 random simulations. Finally, we compute the mean and standard deviation of the stacked density profiles in each of the $144$ regions, to compare with the combined catalogue profile. Note that there may remain additional environmental selection effects to be accounted for; however we expect the above considerations to capture the main factors.

	\section{Results}
	\label{sec:results}
	
	We present the results for our combined catalogue, obtained from the set of $20$ posterior samples previously described. The resulting combined catalogue consists of $150$ voids with radii $>10\Mpch$ within the Local Super-Volume ($135\Mpch$ of the Milky Way). 
	
	A sample of the pure set of anti-halos with high signal-to-noise and which are consistently represented across MCMC samples is given in Table \ref{tab:anti-halos} in Appendix~\ref{app:catalogue}, with the full catalogue available online as supplementary material. We also display the locations of the most robust of these anti-halos on the sky in Fig.~\ref{fig:skyplot}: the outlines show alpha-shapes~\citep{alphashapes} around the particles corresponding to each anti-halo when projected onto the sky in a representative MCMC sample from the 20 used to construct the catalogue. Outlines are broadly similar in the other samples, but fluctuate at a minor level as matter moves around due to unconstrained modes. Note that nearby anti-halos appear larger due to projection effects.
	
	We performed additional checks on the catalogue, including randomly permuting the order in which catalogues were processed. This led to minimal variation in the voids found. 
	
	\subsection{Anti-halo mass functions and void size function}
	\label{sec:amf}
	
	In the anti-halo model of voids, the abundance of voids is predicted by the halo mass function. For large radius voids ($>10\Mpch$), the effect of crushing of voids by surrounding large scale structure is negligible~\citep{pontzen2016inverted}, and so the expected abundance of voids as a function of anti-halo mass can be given by any model for predicting halo mass functions, such as the Tinker mass function~\citep{Tinker:2008ff}. 
	
	We show the anti-halo mass function for the combined catalogue in Fig.~\ref{fig:mass_functions}, finding that it is consistent with the $\Lambda$CDM predictions from the Tinker mass function within the Local Super-Volume. Looking further out to $300\Mpch$, we see that the completeness of the catalogue is much lower, an effect primarily driven by the low signal to noise further from the Milky Way. Since voids are commonly presented in the literature in the form of the void size function (VSF), we also show the abundance of voids within the Local Super-Volume as a function of their radius in Fig.~\ref{fig:vsf}. 
	
	To compute the uncertainty for the bin counts in each mass or radius bin, we treat the counts as a weighted histogram, with weights given by the probability that the mean lies in each bin. We estimate the probability, $p_{ij}$, that void $i$ lies in bin $j$ by bootstrapping the posterior samples to approximate the distribution of the mean mass or radius. Since a void can be either in or out of a bin, the contribution of each void to a bin is a Bernoulli-distributed variable (1 with probability $p_{ij}$, 0 with probability $1-p_{ij}$). The mean bin count, $n_{j}$, and its variance then follow from the sum of $N$ Bernoulli-distributed variables:
	\begin{align}
		n_j =& \sum_{i=1}^{N}p_{ij}, \label{eq:bin_count} \\
		\mathrm{var}(n_j) = & \sum_{i=1}^{N} p_{ij}(1 - p_{ij}), \label{eq:bin_variance}
	\end{align}
	where $N$ is the total number of voids in the catalogue (150). For large numbers of voids in a bin, $n_j$ is approximately Gaussian. However, for the purposes of computing an accurate error bar in low-count bins we estimate the $68\%$ interval for bin counts in Figs.~\ref{fig:mass_functions} and~\ref{fig:vsf} using Monte-Carlo realisations in which each void occupies bin $j$ with probability $p_{ij}$.

	\subsection{Void density profiles}
	
	We present the void density profiles computed using the procedure outlined in Sec.~\ref{sec:void_profiles}. Without any constraints on the voids in random $\Lambda$CDM simulations, the MCMC profiles appear to be outside the $68\%$ interval of profiles (Fig.~\ref{fig:profiles}, left hand panel). However, this is driven in large part by the under-density of the Local Super-Volume. Selecting only $144$ regions of radius $135\Mpch$ whose density matches that of the Local Super-Volume brings the profiles into better agreement by lowering the profile at large distances from the void centre, illustrating the influence of a void's environment on its density profiles. Even better agreement is obtained (Fig.~\ref{fig:profiles}, right hand panel) when we sample a subset of voids from each region whose distribution of central and average densities match that of the Local Super-Volume, indicating consistency with $\Lambda$CDM expectations for voids in the Local Super-Volume.

	\section{Discussion}
	\label{sec:discussion}
	
	The abundance of voids, as shown in Fig.~\ref{fig:mass_functions} and~\ref{fig:vsf}, is consistent with $\Lambda$CDM within $135\Mpch$. Further out, fewer voids are identified (Fig.~\ref{fig:mass_functions}, right panel), but this is consistent with expectations from signal-to-noise considerations: fewer higher signal-to-noise voids are available further out to be matched by the catalogue combination algorithm, and those that are found have lower reproducibility score. Our approach thus correctly excludes random, coincidental alignments of voids which are driven by the prior, rather than the data. 
 
 	This illustrates that our approach by design creates a very pure catalogue, which can be used for high-precision tests of cosmology. The completeness of the catalogue at higher redshifts is primarily limited by the availability of high tracer-density, high-precision redshift data. Therefore, there is broad scope to build larger catalogues of high-quality voids, by including Sloan Digital Sky Survey~\citep[SDSS,][]{abazajian2009seventh} data, and data from upcoming surveys such as {\it Euclid}~\citep{laureijs2011euclid,scaramella2022euclid}; the Vera Rubin Observatory's Legacy Survey of Space and Time~\citep[LSST,][]{2019ApJ...873..111I}; the Roman Space Telescope's High Latitude Survey~\citep{2022MNRAS.512.5311W}; the Dark Energy Spectroscopic Instrument~\citep[DESI,][]{2022AJ....164..207D}; and the Spectro-Photometer for the History of the Universe, Epoch of Reionization, and Ices Explorer~\citep[SPHEREx,][]{2014arXiv1412.4872D}. Fully exploiting the mix of photometric and spectroscopic data in next-generation surveys will however require careful analysis to account for differences in tracer density and redshift precision~\citep{jasche2012bayesian}. The 2M++ results we present in this work, however, already build a robust picture of the voids in the Local Super-Volume.

 	Although we assume a $\Lambda$CDM prior with fixed cosmological parameters in the inference (Sec.~\ref{subsec:posterior_resim}),  the high signal-to-noise regions are demonstrably data-driven. Direct evidence of this for the central $135\Mpch$ region can be seen in the left panel of Fig.~\ref{fig:mass_functions}: by construction, $99\%$ of prior-driven voids would be discarded by our reproducibility score cut, but the abundance of voids we actually see is consistent with $\Lambda$CDM. Where the signal-to-noise ratio drops (beyond $135\Mpch$), our catalogue strongly favours incompleteness over returning spurious prior-driven voids (right panel of Fig.~\ref{fig:mass_functions}).
 	
 	 We expect our catalogue to be insensitive to the $\Lambda$CDM prior used in the inference. BORG seeks to reproduce the final density field, and therefore the void centres and radii are strongly constrained by data. Adopting a different set of cosmological parameters would primarily change the power spectrum, while the locations of void regions are primarily determined by phase information. Such information has previously been shown to be insensitive to cosmological assumptions~\citep{Villaescusa-Navarro_2020,2022A&A...657L..17K}. However, detailed features such as the density profiles may show stronger cosmological dependencies. It would therefore be of clear interest to explore the impact of sampling over cosmological parameters or models~\citep{2022MNRAS.509.3194P,10.1093/mnras/stad432}. A full investigation will be the subject of future work.

	\subsection{Influence of the environment on void density profiles}
	\label{eq:environment}

 It is clear from our results (Fig.~\ref{fig:profiles}) that environmental selection impacts the shape of the (stacked) void density profile. The primary factor driving these results is the fact that the Local Super-Volume density is lower than the cosmological average at the $68\%$ level ($\delta_{\mathrm{MAP}} = -0.041\pm 0.001$, bootstrap error). Accounting for this environmental effect explains the low value of the profile compared to a $\Lambda$CDM `average' profile, and the finer details of the profile are consistent with $\Lambda$CDM once the distribution of central and average void densities in the catalogue is also accounted for. 
	
	Previous studies of voids have indicated a universal void density profile that is found to apply across a wide array of void sizes~\citep{hamaus2014universal,Nadathur:2014qja} and redshifts~\citep{nadathur2014universal}. This remains true of anti-halo void stacks when averaging over all environments, but our results highlight the importance of the environment of cosmic voids in shaping their density profiles. This manifests foremost in the large-radius profile returning to a lower background density (see Fig.~\ref{fig:profiles}, right-hand panel). However, we also see dependence on the profile on the average density of voids. This highlights that in addition to information about gravitational evolution, void profiles contain information about the larger-scale environment in which a void is found. Conversely, this indicates that the environment in which voids are found must be self-consistently accounted for when comparing void catalogues with cosmological models.

	\subsection{Comparison to watershed void catalogues}
	\label{sec:comparison}
	
   \citet{leclercq2015dark} built a watershed void catalogue for the SDSS using \codefont{BORG} and the void-finding code \codefont{VIDE}~\citep{Sutter:2014haa} and \citet{2017JCAP...06..049L} examined more general cosmic web classifiers, including voids. \citet{leclercq2015dark} also performed simulations of posterior samples, an approach they refer to as non-linear filtering, which shares similarities to the posterior resimulation approach we use in this work, but lacks an object-by-object matching. As such, while they were able to estimate aggregate properties of their void catalogue with a Blackwell-Rao approach, it was not possible to study the reliability of individual voids. By contrast, posterior resimulation requires satisfying accuracy requirements on field level inference (S23), and an assessment of the reliability of the structures appearing in resimulated posterior samples (this work).
	
	There is an effective limit of $\sim 25\mathrm{-}30\Mpch$ on the radius an anti-halo void can have, due to the absence of significant number of halos above $10^{15}M_{\odot}h^{-1}$. Higher radius voids are frequently found in other void catalogues such as \codefont{GIGANTES}~\citep{kreisch2022gigantes}, and other catalogues based on the BOSS DR12 data such as the watershed catalogues by~\citet{Nadathur:2016nqr,mao2017cosmic}. However, these typically represent shallower, large-volume voids which in the anti-halo picture would not yet have virialised. Particles within such watershed voids are typically found within smaller anti-halos in analogy to the bottom-up picture of structure formation with halos - see ~\citet{Pontzen:2015eoh} for example. 
	
	\subsection{Concluding remarks}
	
	We have presented a void catalogue for the Local Super-Volume, constructed by robustly combining anti-halo voids identified in posterior resimulations of the dark matter density field using initial conditions obtained with field-level inference. The methodology used to construct the catalogue is general and can be applied to posterior resimulations of field-level inferred initial conditions for any galaxy catalogue.
	
	Our catalogue is complete out to $135\Mpch$, identifying all anti-halos with radii $>10\Mpch$. We have shown that it is compatible with $\Lambda$CDM expectations, both in the abundance of voids, and in the shape of the stacked void density profiles. One application of the catalogue, which we leave to future work, is to constrain cosmological effects whose impact is strongly felt in low density regions. An important example is neutrino masses, which are known to impact the abundance of voids~\citep{10.1093/mnras/stz1944}, and their density profiles~\citep{massara2015voids}. Additionally, modified gravity models such as $f(R)$~\citep{hu2007models} and nDGP~\citep{dvali2000metastable} theories are known to affect the shapes of voids and their density profiles~\citep{falck2018using, cai2014testing,cautun2018santiago,paillas2019santiago}. While such theories are heavily constrained in their ability to explain dark energy~\citep{saridakis2021modified}, cosmological constraints are still capable of constraining the properties of gravity itself~\citep{shankaranarayanan2022modified}. A challenge which remains, however, is how to disentangle the effects of modified gravity from that of massive neutrinos, since both have similar effects on void density profiles~\citep{baldi2014cosmic,hagstotz2019breaking}. 
 
 In future work, we will explore the impact on the detailed properties of voids within the catalogue, if one assumes different cosmologies within field-level inference. These developments will be crucial for obtaining robust constraints on gravity, cosmology and particle physics using the Local Super-Volume. In the meantime, however, we will test the consistency of the local void population with $\Lambda$CDM predictions, an important first step enabled by the catalogue of voids presented in this work.

    \section*{Acknowledgements}
    We thank Metin Ata, Harry Desmond, Stuart McAlpine, and Daniel Mortlock for useful discussions regarding this work. This project has received funding from the European Research Council (ERC) under the European Union’s Horizon 2020 research and innovation programmes (grant agreement no. 101018897 CosmicExplorer and 818085 GMGalaxies). This work has been enabled by support from the research project grant ‘Understanding the Dynamic Universe’ funded by the Knut and Alice Wallenberg Foundation under Dnr KAW 2018.0067. SS and HVP are additionally supported by the Göran Gustafsson Foundation for Research in Natural Sciences and Medicine. HVP and JJ acknowledge the hospitality of the Aspen Center for Physics, which is supported by National Science Foundation grant PHY-1607611. The participation of HVP and JJ at the Aspen Center for Physics was supported by the Simons Foundation. JJ acknowledges support by the Swedish Research Council (VR) under the project 2020-05143 -- ``Deciphering the Dynamics of Cosmic Structure" and by the Simons Collaboration on “Learning the Universe”. This work was partially enabled by the UCL Cosmoparticle Initiative. The computations/data handling were enabled by resources provided by the Swedish National Infrastructure for Computing (SNIC) at Link\"{o}ping University, partially funded by the Swedish Research Council through grant agreement no. 2018-05973. This research also utilized the Sunrise HPC facility supported by the Technical Division at the Department of Physics, Stockholm University. This work was carried out with support from the Aquila Consortium\footnote{\url{https://aquila-consortium.org}}.
	
	\section*{Author contributions}

    We outline the different contributions below using key-words based on the CRediT (Contribution Roles Taxonomy) system. 

    {\bf SS}: data curation; investigation; formal analysis; software; visualisation; writing -- original draft preparation. 
    
    {\bf HVP}: conceptualisation; methodology; validation and interpretation; writing (original draft; review and editing); funding acquisition.
    
    {\bf AP}: conceptualisation; methodology; validation and interpretation; writing (original draft; review and editing). 
    
    {\bf JJ}: methodology; data curation; resources; software; validation and interpretation; writing – review 
    
    {\bf GL}: data curation; resources; software; writing -- review. 
	
	\section*{Data availability}
	The data underlying this article will be shared on reasonable request to the corresponding author. The full combined catalogue in Table~\ref{tab:anti-halos} is supplied as supplementary material in csv format. Additional data, including the anti-halo catalogues for each MCMC sample, is available on Zenodo at \url{https://doi.org/10.5281/zenodo.10160612}.

    \interlinepenalty=10000
    
    \bibliographystyle{mnras}
    \bibliography{borg_antihalos_paper.bib}

	\appendix

	\section{The anti-halo catalogue}
	\label{app:catalogue}	
	
	In Table~\ref{tab:anti-halos}, we present the final catalogue after applying the combination algorithm in Sec.~\ref{sec:construction}. Uncertainties are computed as the standard deviation of the mean over all anti-halos matching to a given void (which may not include all samples if a void is missing in some samples). The voids are sorted by  the reproducibility score, which quantifies how robustly a void appears in all catalogues (a reproducibility score of 1 indicates that a void appeared in all catalogues). As discussed in Sec.~\ref{sec:construction}, voids with reproducibility scores not significantly higher than would be found by chance alone are not included. We also show the reproducibility score thresholds for each radius bin in Table~\ref{tab:thresholds}.
	
	\begin{table*}
		\centering
		\begin{tabular}{c|c|c|c|c|c|c|c|c}
			\hline
			ID & Radius (\(h^{-1}\mathrm{Mpc}\)) & Mass (\(10^{14}h^{-1}M_{\odot}\)) & R.A. (deg.) & Dec. (deg.) & z & Distance (\(h^{-1}\mathrm{Mpc}\)) & SNR & Reproducibility Score \\
			\hline
			\begin{filecontents*}{void_catalogue.csv}
ID,Radius $(h^{-1}\mathrm{Mpc})$,Radius uncertainty $(h^{-1}\mathrm{Mpc})$,Mass $(10^{14}h^{-1}M_{\odot})$,Mass uncertainty $(10^{14}h^{-1}M_{\odot})$,R.A. (deg.),Dec. (deg.),z,Distance $(h^{-1}\mathrm{Mpc})$,SNR,Reproducibility Score,
1,21,0.3,9.5,0.4,343,37.2,0.026,113,18.3,1,
2,14,0.2,3,0.2,0.263,-39.9,0.01,45.1,147,1,
3,17,0.3,5.1,0.2,139,8.68,0.031,135,21.3,1,
4,14,0.2,3.2,0.2,40.6,-50.2,0.0072,32,23.7,1,
5,11,0.2,1.4,0.07,218,-25.6,0.021,93.7,79.2,1,
6,13,0.1,2.5,0.09,223,24.1,0.018,79.2,147,1,
7,13,0.1,2,0.07,49.6,14.5,0.011,48,42.2,0.95,
8,11,0.2,1.6,0.08,318,-19.7,0.0097,42.7,35,0.95,
9,10,0.2,1.2,0.09,212,-14.4,0.017,74.3,44.6,0.95,
10,11,0.2,1.5,0.06,174,-28.6,0.012,52.4,30,0.95,
11,14,0.4,2.9,0.3,246,-17.5,0.024,106,73.3,0.95,
12,17,0.3,5.4,0.3,287,-37.9,0.025,109,80,0.95,
13,12,0.3,2.1,0.2,253,29.4,0.017,76.5,52.2,0.9,
14,11,0.2,1.5,0.07,166,37.6,0.011,48.5,122,0.9,
15,12,0.2,2,0.1,115,24.2,0.024,108,19.4,0.9,
16,11,0.2,1.4,0.09,232,14.8,0.027,120,35.5,0.9,
17,12,0.2,1.8,0.1,137,1.16,0.023,101,153,0.85,
18,10,0.3,1.2,0.1,265,46.3,0.0085,37.5,161,0.85,
19,14,0.4,3.1,0.3,50.7,-25.9,0.02,86.1,91.2,0.85,
20,10,0.3,1.3,0.1,54.2,-36.7,0.023,102,63.5,0.85,
21,16,0.3,4.2,0.3,22.4,-51.9,0.025,111,13.9,0.85,
22,12,0.3,1.7,0.1,244,58.1,0.017,76.3,23.6,0.85,
23,10,0.1,1.1,0.06,178,42.8,0.019,83.4,58.5,0.85,
24,13,0.2,2.3,0.1,20.5,29.1,0.019,84.8,12.1,0.8,
25,15,0.4,3.9,0.3,125,9.6,0.016,72.6,20.8,0.8,
26,13,0.4,2.4,0.2,221,-52.5,0.017,75.6,30.5,0.8,
27,15,0.3,3.3,0.3,217,6.09,0.029,127,22,0.75,
28,12,0.4,1.9,0.2,355,-43.7,0.016,71,36.5,0.75,
29,21,0.6,10,0.9,337,17,0.03,134,22.9,0.75,
30,10,0.2,1,0.07,243,-31.4,0.022,96.2,50.7,0.75,
31,10,0.3,1.2,0.1,291,61.6,0.012,51.5,83.1,0.75,
32,12,0.2,1.6,0.1,229,-20.7,0.022,96.7,41,0.75,
33,11,0.2,1.4,0.08,205,-6.75,0.023,99.5,44.5,0.75,
34,11,0.3,1.3,0.09,37.4,30.6,0.017,72.9,18,0.75,
35,12,0.4,2,0.2,344,-13.1,0.027,117,26.7,0.75,
36,13,0.3,2.1,0.2,53.3,29.6,0.021,92.1,82.3,0.75,
37,17,0.5,5,0.5,329,28.7,0.021,90.5,14.9,0.75,
38,14,0.6,2.9,0.4,128,12.1,0.026,113,85.3,0.7,
39,10,0.3,1.2,0.1,197,51,0.03,132,39.2,0.7,
40,17,0.3,5.7,0.4,227,32.8,0.027,121,14.1,0.7,
41,11,0.4,1.5,0.2,336,-30.3,0.029,128,33.7,0.7,
42,15,0.5,3.8,0.4,59.5,27.7,0.026,115,26.4,0.7,
43,11,0.4,1.3,0.2,296,-5.42,0.0095,41.9,21.9,0.7,
44,10,0.4,1.1,0.1,249,16.2,0.028,121,50,0.65,
45,11,0.3,1.5,0.2,240,33.3,0.026,113,24.7,0.65,
46,13,0.4,2.4,0.2,15.8,-44.1,0.027,120,13.1,0.65,
47,12,0.5,2,0.3,5.56,41.9,0.017,76.1,34,0.65,
48,11,0.3,1.3,0.2,160,-33.9,0.028,122,93.4,0.65,
49,12,0.5,2,0.3,129,29.5,0.024,106,39.6,0.65,
50,11,0.4,1.5,0.2,344,-37.6,0.015,65.7,106,0.65,
51,12,0.3,1.6,0.1,342,-44.1,0.026,116,24.9,0.6,
52,12,0.4,1.9,0.2,308,-9.08,0.027,117,138,0.6,
53,12,0.3,1.8,0.2,357,16.8,0.021,92.2,23.8,0.6,
54,10,0.3,1.1,0.1,212,-6.38,0.013,56.3,33.8,0.6,
55,13,0.4,2.3,0.2,210,23.1,0.029,129,13.5,0.6,
56,11,0.3,1.5,0.2,113,10.2,0.027,117,47.4,0.6,
57,11,0.4,1.7,0.2,192,-4.76,0.022,97.4,76.5,0.6,
58,12,0.3,1.5,0.1,27.4,23.4,0.015,65.3,15.6,0.6,
59,15,0.5,3.4,0.3,347,-35.8,0.031,134,14.1,0.6,
60,11,0.2,1.4,0.08,151,-19.9,0.027,120,30.3,0.6,
61,10,0.3,1.2,0.1,175,1.86,0.024,106,15.1,0.6,
62,16,0.6,4.9,0.5,6.72,-19.7,0.028,122,20.6,0.6,
63,12,0.5,1.8,0.2,351,-49.7,0.026,116,25.7,0.6,
64,11,0.3,1.6,0.1,349,-24.2,0.025,110,46.1,0.55,
65,12,0.3,1.9,0.1,28.5,6.14,0.024,107,56,0.55,
66,12,0.3,1.7,0.1,354,-11.7,0.025,111,13.7,0.55,
67,14,0.5,2.7,0.3,236,-8.68,0.029,127,15,0.55,
68,10,0.4,1.2,0.1,42.4,4.74,0.023,102,25.8,0.55,
69,11,0.4,1.5,0.2,327,13.8,0.026,112,21.6,0.5,
70,10,0.3,1.2,0.1,95.6,36.1,0.025,109,111,0.5,
71,12,0.5,2,0.3,24.9,-35.8,0.029,126,28.4,0.5,
72,10,0.4,1.1,0.1,11.8,40.5,0.019,82.1,35.1,0.5,
73,12,0.4,2.1,0.2,195,-27.2,0.025,112,12.5,0.5,
74,10,0.2,1.1,0.1,174,51.2,0.028,123,87.4,0.5,
75,10,0.3,1.3,0.1,98.6,-26.7,0.022,98.2,99.4,0.5,
76,13,0.6,2.2,0.3,72.8,-2.3,0.027,118,27.5,0.5,
77,15,0.6,3.7,0.4,259,56.9,0.026,116,20.4,0.45,
78,10,0.5,1.2,0.1,111,14.7,0.016,69.8,18.9,0.45,
79,13,0.4,2.2,0.2,297,-31.7,0.03,133,24.4,0.45,
80,11,0.3,1.2,0.1,321,-7.9,0.023,102,49,0.45,
81,11,0.4,1.3,0.1,182,29.4,0.025,112,16.6,0.45,
82,11,0.6,1.5,0.3,160,-21.9,0.03,133,28.3,0.45,
83,13,0.4,2.5,0.2,44.3,24,0.027,120,35.4,0.4,
84,11,0.3,1.2,0.1,51.2,19.7,0.006,26.7,12.4,0.4,
85,11,0.5,1.3,0.2,139,-7.73,0.024,107,80.2,0.4,
86,12,0.4,1.9,0.2,208,7.73,0.03,133,31.1,0.4,
87,13,0.4,2.1,0.2,329,-3.42,0.027,117,43.3,0.4,
88,10,0.5,1.2,0.2,99.8,-24.6,0.023,102,75.2,0.4,
89,11,0.4,1.6,0.1,74,-35.7,0.017,74.9,124,0.4,
90,11,0.5,1.3,0.2,110,33.9,0.022,95.4,25.4,0.4,
91,12,0.5,1.8,0.2,257,3.7,0.025,112,37.5,0.4,
92,11,0.3,1.2,0.1,171,42.9,0.027,119,40.9,0.4,
93,11,0.4,1.5,0.2,191,-0.205,0.021,91.1,102,0.4,
94,13,0.5,2.2,0.3,133,15.3,0.029,127,24.2,0.4,
95,11,0.6,1.7,0.2,85.9,71.5,0.025,112,13.3,0.4,
96,11,0.5,1.5,0.2,102,19.1,0.022,98.3,15.8,0.35,
97,10,0.2,1.1,0.09,351,39.3,0.019,82.3,16.5,0.35,
98,11,0.4,1.5,0.2,52.3,55.8,0.016,70.6,48.9,0.35,
99,10,0.5,1.2,0.2,45.9,41.7,0.03,130,12.5,0.35,
100,12,0.4,2.1,0.2,112,-42.8,0.024,106,23.5,0.35,
101,14,0.7,2.7,0.4,202,-3.53,0.027,121,14.2,0.35,
102,10,0.4,1.3,0.2,300,-70.9,0.019,83.4,26.6,0.35,
103,11,0.7,1.5,0.3,233,36.3,0.016,72.3,91.7,0.35,
104,14,0.5,2.7,0.2,207,-2.03,0.028,123,13,0.35,
105,12,0.5,1.8,0.3,284,28.1,0.025,112,11.6,0.3,
106,11,0.6,1.4,0.3,341,-9.06,0.026,113,27.9,0.3,
107,10,0.2,1.1,0.05,300,11.3,0.027,117,25.6,0.3,
108,10,0.6,1.2,0.2,306,14.8,0.027,120,74.5,0.3,
109,12,0.5,1.9,0.3,9.8,46.2,0.0073,32.2,21.4,0.3,
110,13,0.3,2.6,0.3,39,-19.7,0.029,130,21.2,0.3,
111,10,0.6,1.1,0.2,72.6,-4.96,0.027,117,30.8,0.3,
112,14,0.6,2.8,0.4,68.3,6.49,0.028,121,18.1,0.3,
113,11,0.5,1.2,0.2,282,-36.3,0.028,123,25.3,0.3,
114,11,0.4,1.4,0.1,108,33.3,0.026,115,75.7,0.3,
115,14,0.5,2.9,0.3,133,11.3,0.026,115,76.6,0.3,
116,13,0.7,2.2,0.4,117,12.4,0.017,76.4,16.1,0.3,
117,12,0.6,2,0.3,353,-48.7,0.029,126,48.2,0.3,
118,12,0.3,1.8,0.2,14.2,36.1,0.0059,26.1,19.6,0.3,
119,14,0.9,3,0.6,15.3,9.5,0.017,74.1,14.8,0.3,
120,10,0.3,1.1,0.1,162,-13.6,0.023,101,20.5,0.3,
121,15,0.8,3.2,0.6,69,24.4,0.028,122,26.3,0.25,
122,15,0.8,3.9,0.7,272,29.5,0.022,98.3,30.3,0.25,
123,14,0.4,3.3,0.3,357,44.2,0.023,100,14.9,0.25,
124,13,0.6,2.2,0.3,31,45.4,0.027,117,11.8,0.25,
125,10,0.6,1.2,0.3,326,-4.95,0.028,124,82.7,0.25,
126,12,0.6,1.9,0.2,158,-51.2,0.023,103,54.2,0.25,
127,11,0.6,1.1,0.2,133,17.4,0.027,121,32.1,0.25,
128,11,0.6,1.5,0.2,56.4,-29.3,0.016,71.5,71.7,0.25,
129,11,0.7,1.3,0.3,24.7,43.3,0.027,120,20.4,0.25,
130,11,0.6,1.3,0.2,27.8,14.3,0.026,115,29.2,0.25,
131,11,0.4,1.4,0.1,309,14.4,0.024,106,32.6,0.25,
132,13,0.8,2.1,0.4,137,-3.73,0.025,109,55.5,0.25,
133,12,0.5,1.7,0.2,137,36.8,0.024,105,32.5,0.25,
134,10,0.3,1.1,0.1,45.6,30.2,0.02,88.2,88.6,0.25,
135,12,0.6,1.9,0.3,6.52,-18.3,0.025,111,51,0.25,
136,11,0.2,1.3,0.1,11.4,-12.8,0.03,130,11.3,0.25,
137,13,0.6,2.2,0.3,318,81,0.029,127,19.4,0.25,
138,11,0.4,1.3,0.1,273,23.6,0.03,132,220,0.25,
139,13,0.5,2.1,0.3,287,37.5,0.022,96.3,12.8,0.25,
140,10,0.4,1.1,0.1,29.2,40.6,0.03,130,19.1,0.25,
141,10,0.5,1.2,0.1,6.75,28.8,0.026,116,22.8,0.25,
142,12,0.5,1.7,0.2,344,-4.35,0.025,109,13.4,0.25,
143,11,0.5,1.3,0.2,83.5,-41.8,0.013,55.6,134,0.25,
144,10,0.5,1.3,0.2,57.1,52.1,0.018,78.2,41.8,0.25,
145,15,1,4.3,1,295,4.81,0.025,110,21,0.2,
146,15,0.9,3.7,0.7,293,8.44,0.023,103,22.3,0.2,
147,13,1,2.5,0.5,291,-9.52,0.0092,40.9,21.4,0.2,
148,14,0.9,3.2,0.6,255,7.34,0.027,120,48.4,0.2,
149,19,0.9,7.9,1,64.6,27.6,0.027,119,20.2,0.2,
150,14,0.6,3,0.3,293,10.8,0.021,92.8,18.9,0.2,
\end{filecontents*}
\csvreader[column count=11,filter expr={test{\ifnumgreater{\value{csvinputline}}{1}} and test{\ifnumless{\value{csvinputline}}{12}}},no head,late after line=\\]{void_catalogue.csv}{1=\one,2=\two,3=\three,4=\four,5=\five,6=\six,7=\seven,8=\eight,9=\nine,10=\ten,11=\eleven}{$\one$ & $\two\pm\three$ & $\four\pm\five$ & $\six$ & $\seven$ & $\eight$ & $\nine$ & $\ten$ & $\eleven$}
			\multicolumn{9}{c}{\ldots Continued in supplementary materials.}\\
			\hline
		\end{tabular}
		\caption{\label{tab:anti-halos} First $10$ of $150$-total voids in the anti-halo catalogue for voids within $135\Mpch$. Radii and mass uncertainties are computed as the standard error on the mean over the anti-halos in each MCMC sample matching on to a given void. The signal-to-noise ratio (SNR) is the mean signal to noise for each of the same contributing anti-halos, computed as the average of $\delta^2/\sigma_{\delta}^2$ as discussed in Sec.~\ref{sec:construction}. The voids are sorted by reproducibility score, defined as the fraction of MCMC samples which contain a representative anti-halo for a given void. The entire catalogue is provided as supplementary material.}
	\end{table*}

	\begin{table*}
		\centering
		\begin{tabular}{c|c}
			\hline
			Radius Bin Range ($\Mpch$) & Reproducibility Score Threshold \\
			\hline
			$10$--$11.4$ & 0.2 \\
			$11.4$--$12.9$ & 0.2 \\
			$12.9$--$14.3$ & 0.15 \\
			$14.3$--$15.7$ & 0.15 \\
			$15.7$--$17.1$ & 0.15 \\
			$17.1$--$18.6$ & 0.142 \\
			$18.6$--$20$ & 0.1 \\
			\hline
		\end{tabular}
		\caption{\label{tab:thresholds} Reproducibility score thresholds in each radius bin, above which 99\% of voids in $\Lambda$CDM simulations are excluded.}
	\end{table*}

\label{lastpage}
	
\end{document}